\documentclass{aa}  
\usepackage{graphicx}
\usepackage{color}
\usepackage{txfonts}
\usepackage{natbib}
\usepackage{hyperref}
\usepackage{caption}
\usepackage{subcaption}

\usepackage[switch, modulo]{lineno}

\begin{document} 

\title{On the use of the Fourier Transform to determine the projected rotational
  velocity of line-profile variable B stars\thanks{The spectroscopic time series
    used in this paper, along with many others, are available from {\tt
      http://newton.ster.kuleuven.be/$\sim$roy/helas/}}}

\author{C. Aerts\inst{1,2} \and S.\ S\'{\i}mon-D\'{\i}az\inst{3,4} \and
  P.\ J.\ Groot\inst{2} 
\and P.\ Degroote\inst{1,}\thanks{Postdoctoral Fellow of the
    Fund for Scientific Research of Flanders (FWO), Belgium} }

\institute{Instituut voor Sterrenkunde, KU Leuven, Celestijnenlaan 200D, 3001
  Leuven, Belgium\\ \email{Conny.Aerts@ster.kuleuven.be}
           \and
Department of Astrophysics/IMAPP, Radboud University Nijmegen,
             6500 GL Nijmegen, The Netherlands
           \and
Instituto de Astrof\'{\i}sica de Canarias, 38200, La Laguna, Tenerife, Spain
           \and
Departamento de Astrof\'{\i}sica, Universidad de La Laguna, 38205, La Laguna,
Tenerife, Spain }

\date{Received ; Accepted}
   
\titlerunning{The Fourier Transform applied to line-profile variable B stars}
\authorrunning{C.\ Aerts et al.} 

 
\abstract {The Fourier Transform method is a popular tool to derive the
  rotational velocities of stars from their spectral line profiles.  However,
  its domain of validity does not include line-profile variables with
  time-dependent profiles.}  {We investigate the performance of the method for
  such cases, by interpreting the line-profile variations of spotted { B} stars, and
  of pulsating {B} tars, as if their spectral lines were caused by uniform surface
  rotation along with macroturbulence.} {We perform time-series analysis and
  harmonic least-squares fitting of various line diagnostics and of the outcome
  of several implementations of the Fourier Transform method.}  {We find that
  the projected rotational velocities derived from the Fourier Transform vary
  appreciably during the pulsation cycle whenever the pulsational and rotational
  velocity fields are of similar magnitude. The macroturbulent velocities
  derived while ignoring the pulsations can vary with tens of km\,s$^{-1}$
  during the pulsation cycle. The temporal behaviour of the deduced rotational
  and macroturbulent velocities are in antiphase with each other. The
  rotational velocity is in phase with the second moment of the line profiles.}
{The application of the Fourier method to stars with considerable pulsational
  line broadening may lead to an appreciable spread in the values of the
  rotation velocity, and, by implication, of the deduced value of the
  macroturbulence. These two quantities should therefore not be derived from
  single snapshot spectra if the aim is to use them as a solid diagnostic for
  the evaluation of stellar evolution models of slow to moderate rotators. }


\maketitle


\section{Introduction}

Along with the effective temperature, gravity, and chemical composition of the
stellar surface, the rotation velocity of stars has been considered as an
important observational diagnostic to evaluate stellar evolution models,
particularly in the case of massive stars \citep[e.g.,][for an extensive
  monograph on the theory of stellar evolution including rotation]{Maeder2009}.
Unfortunately, the quantity which is usually measured from high-resolution high
signal-to-noise spectroscopy is the projected rotation velocity $v\sin i$
instead of the equatorial rotation velocity. Even in the case when the rotation
frequency is available with high precision from time series analysis of
spectro-polarimetry or from asteroseismology \citep[e.g.][for a compilation of
  massive stars with these quantities available]{Aerts2014}, the stellar radius
is not within reach of direct observations for large samples of stars. Hence,
the outcome of stellar evolution computations is often compared with measured
values of $v\sin i$ \citep[e.g.][and references
  therein]{Hunter2009,Brott2011,Markova2014}. An accurate determination of
$v\sin i$ is thus essential to test compliance with stellar models. Here, we
focus on the determination of $v\sin i$ for slowly to moderately rotating B-type
line-profile variable stars.

Since the development of high-resolution spectroscopy initiated more than three
decades ago, it is well known that rotation is not the only line-broadening
mechanism at work in massive stars. 
{
We refer to
\citet{Smith1978,Campos1980} for early pioneering work in this topic, 
where macroturbulent velocities were introduced to obtain good fits to the
  observed line profiles, while 
\citet{AertsDeCat2003} and 
\citet{Howarth2004} presented 
early reviews in the case of pulsational
line-broadening of main-sequence B stars
and macroturbulent line-broadening of O-type stars and
BA-type supergiants, respectively.
  The physical origin of the often large macroturbulent velocities
adopted to fit spectral lines of hot stars is
  still uncertain and may be quite different for various types of objects.  This
  is in contrast with the situation for cool stars, where the macroturbulent
  velocities necessary to obtain good spectral lines fits are well identified as
  being due to the limitations of using 1D model atmospheres rather than
  considering the convective motions properly in a 3D description
  \citep{Asplund2000}. 
Due to the unknown physical nature of the macroturbulent velocities
  in massive stars, 
}
the separation of the rotational line-broadening from the combination of other
broadening mechanisms is far from trivial, particularly if only a single or a
few snapshot spectra are available. In the case of pulsating B stars, the
rotational velocity can be disentangled from the pulsational and microturbulent
broadening from time-resolved spectroscopy that covers the entire beating cycle
of the pulsations \citep[e.g.,][Chapter 6]{Aerts2010}. Such data sets are rather
scarce and are not available in the case of large sample studies.

With the aim to perform an extensive study of stellar rotation in OB-type stars
covering all evolutionary stages, 
\citet[][hereafter termed
  SDH14]{SimonDiaz2014} 
reviewed the status of the derivation of $v\sin i$, and
along with it of macroturbulence ($v_{\rm macro}$), from high-precision
spectroscopy. Their analyses 
led to the conclusion that adequate values for both
$v\sin i$ and $v_{\rm macro}$ can best be obtained from a line-profile analysis
relying on the application of the Fourier Transform (FT) method \citep[e.g.][and
  references therein]{Carroll1933,Smith1976,Dravins1990,Gray2005}, followed by a
goodnes-of-fit (GOF hereafter) approach to determine $v_{\rm macro}$ using a
radial-tangential approximation for the macroturbulence.  Only in the case where
$v\sin i$ from the FT and from the GOF approaches agree, can one trust the
derived values of these two quantities.

As pointed out in SDH14, and also stressed by \citet{Sundqvist2013}, the
derivation of $v\sin i$ from the first zero of the Fourier transform of a
spectral line may be quite erroneous for low rotation velocities and in the
presence of large macroturbulence, the quoted limit of applicability being
$v\sin i = 50\,$km\,s$^{-1}$
{ in the case where 
the line profile is affected by an important macroturbulent contribution.}
This is particularly the case when the spectral
resolution of the instrument is limited.  The risk of misinterpreting
line-profile broadening in terms of macroturbulence while ignoring pulsational
velocity fields due to gravity modes and their accompanying line broadening was
already stressed by \citet{Aerts2009} from extensive simulations of line-profile
variations. 

With the current work, we shed new light on the matter by focusing on B stars in
the core-hydrogen burning phase and in the regime 
{ $v\sin i < 80\,$km\,s$^{-1}$},
keeping in mind that a large fraction of these stars are line-profile variables
\citep[e.g.][]{Telting2006}, a fact that tends to be ignored when deriving
$v_{\rm macro}$.  We illustrate how various versions of the FT method give rise
to a large range of $v\sin i$ and the deduced $v_{\rm macro}$ values during the
variability cycle for stars whose surface displays inhomogeneities and/or
stellar pulsations. It turns out that the estimation of $v\sin i$ from snapshot
spectra can be subject to considerable uncertainty when dealing with
line-profile variable stars.

\section{Selected line-profile variable B stars}

\begin{figure*}[t!]
\begin{center}
\rotatebox{270}{\resizebox{6cm}{!}{\includegraphics{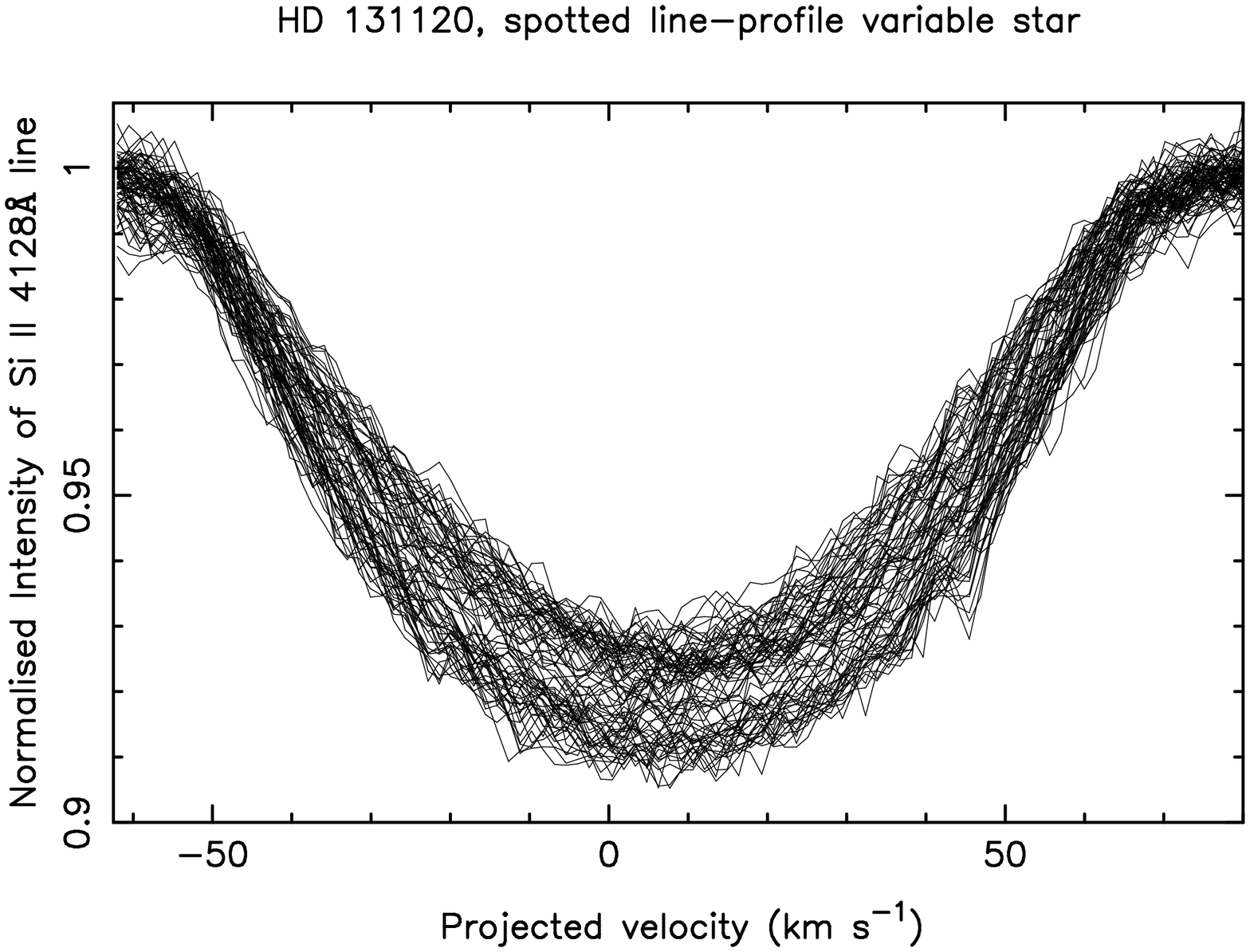}}}\hspace{0.5cm} 
\rotatebox{270}{\resizebox{6cm}{!}{\includegraphics{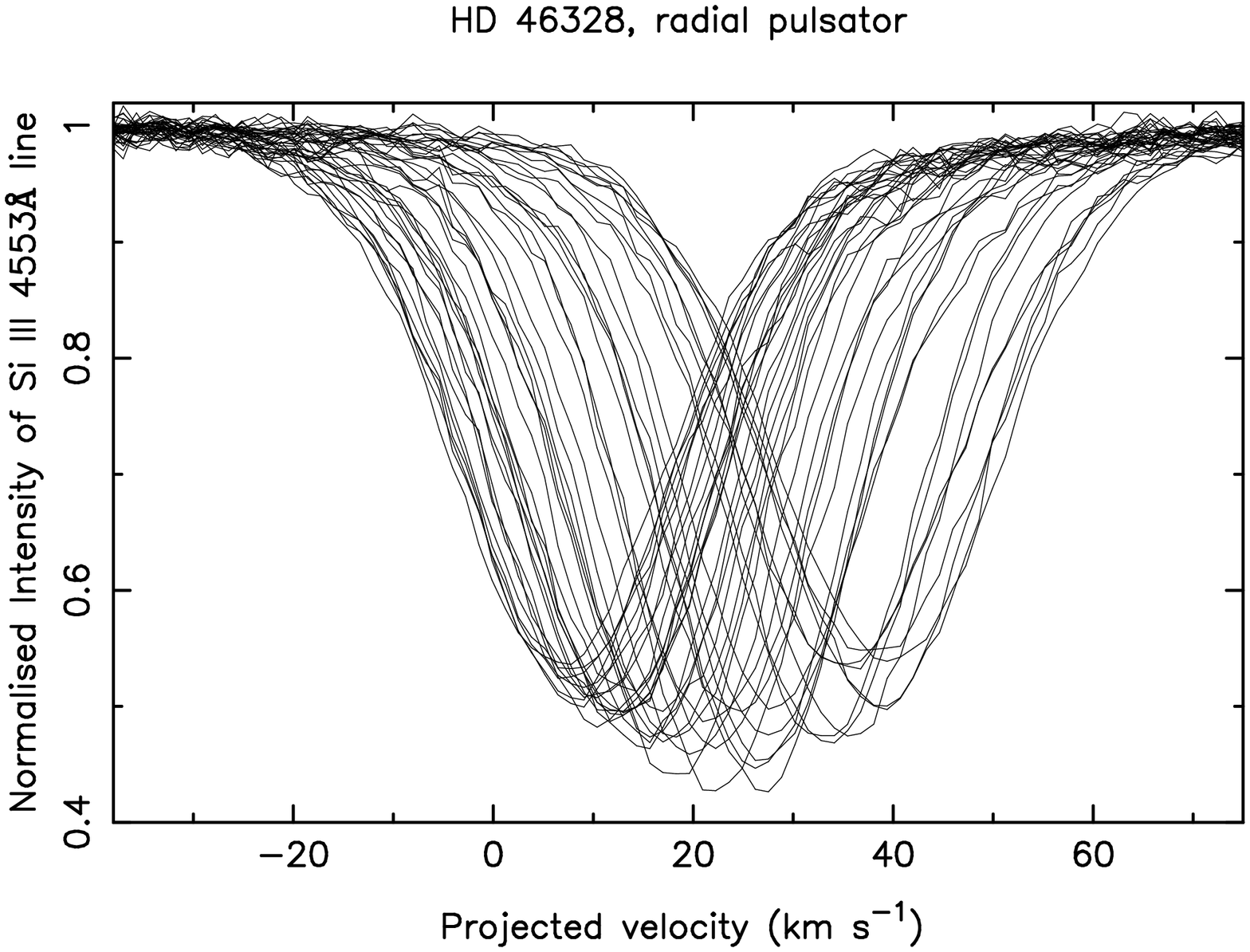}}}\\[0.5cm] 
\rotatebox{270}{\resizebox{6cm}{!}{\includegraphics{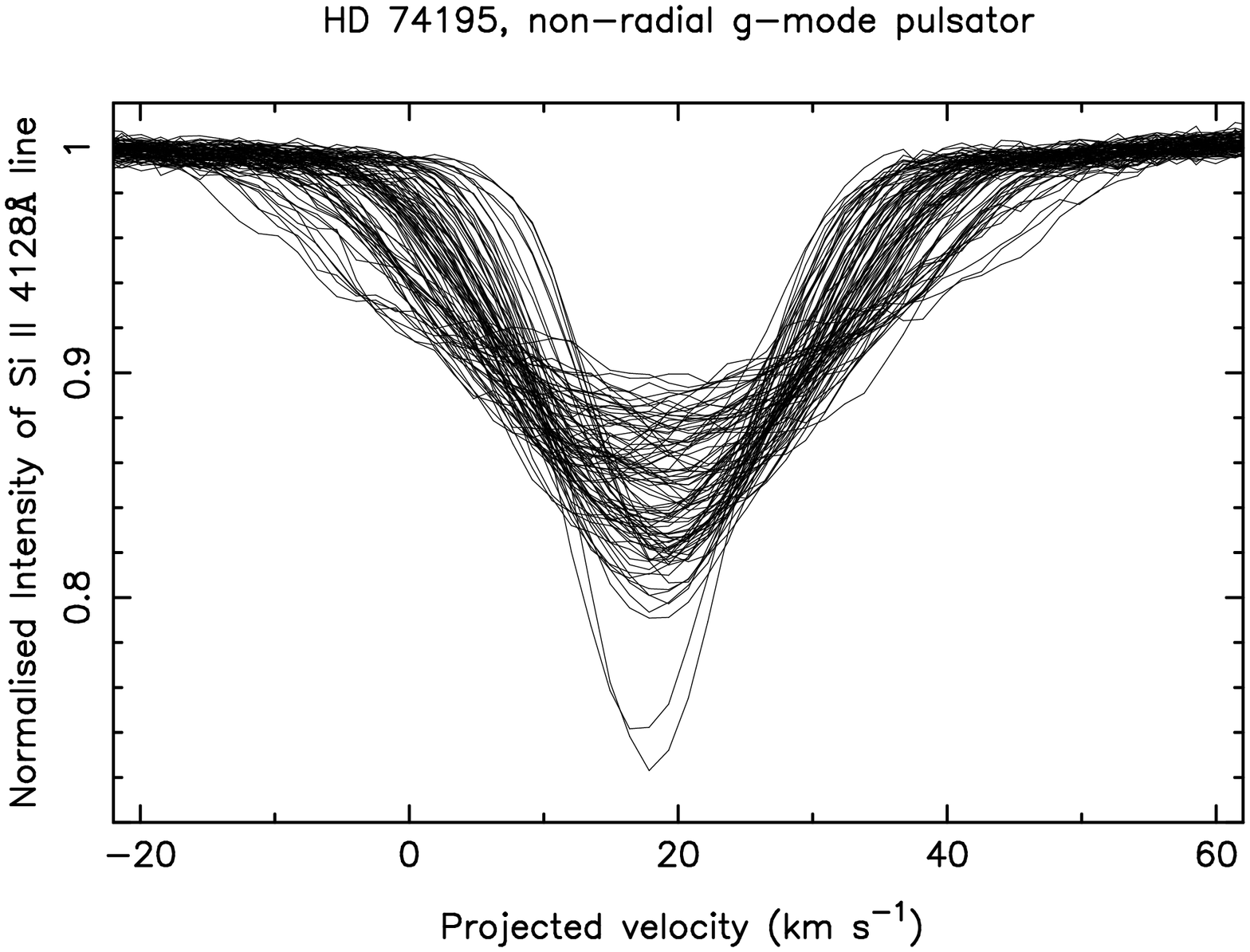}}}\hspace{0.5cm} 
\rotatebox{270}{\resizebox{6cm}{!}{\includegraphics{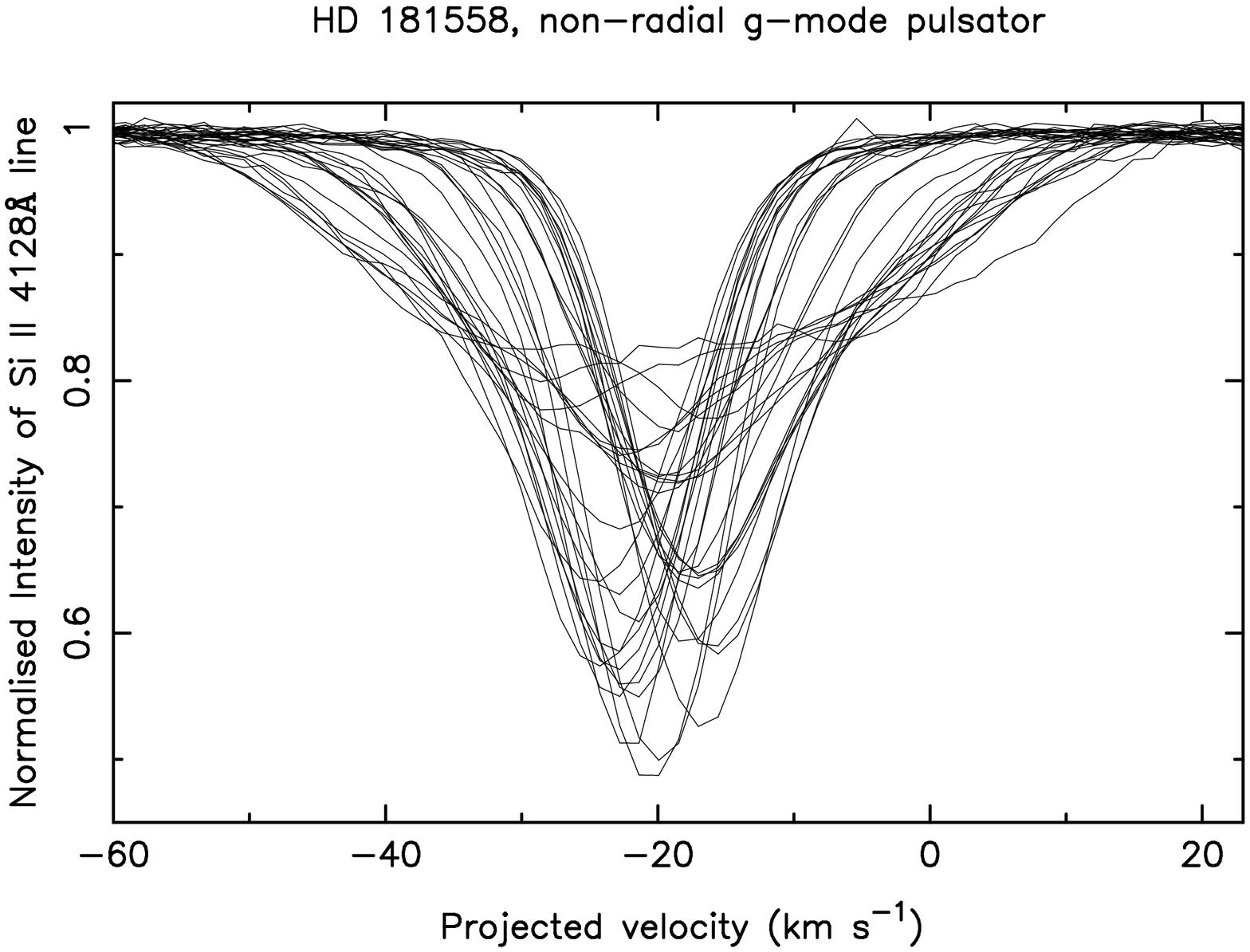}}} 
\end{center}
\caption{Line profiles used in this study for { four} of the targets listed in
  Table\,\ref{table1}. The nature of the variability, the data source, and the
  laboratory wavelength of the spectral line corresponding to the zero point are
  discussed in the text.  For HD\,46328 we only show one tenth of the data for
  visibility reasons.}
\label{LPVs}
\end{figure*}

In this work, we consider a few prototypical well-studied line-profile variable
core-hydrogen burning B stars as the simplest case studies to assess the
robustness of the FT method when dealing with asymmetric line profiles.  In a
future study, a sample of O-type stars and B-type supergiants whose line-profile
properties are yet unknown will be considered (S\'{\i}mon-D\'{\i}az et al., in
preparation).

Line-profile variability of B stars can have several causes. The two most
prominent ones are chemical and/or temperature inhomogeneities at the stellar
surface giving rise to monoperiodic rotational modulation in the line profiles,
or stellar pulsations leading to mono- or multiperiodic profile changes
depending on the character of the excited modes. A mixture of these two, i.e.,
pulsations and surface spots, may also occur
\citep[e.g.][]{Telting1997,Uytterhoeven2005,Briquet2012}.

Depending on their dominant restoring force, the pulsations are either pressure
(p-) modes, gravity (g-) modes, or mixed modes. The p-modes give rise to
dominant radial velocity fields while the g-modes lead to dominant tangential
velocities. We point out that the pulsations also give rise to temperature
variations across the stellar surface \citep[e.g.][]{DeRidder2002}, but that
these are only a minor secondary effect compared to the line-profile variability
induced by the pulsational velocity field, even in the case of
temperature-sensitive spectral lines \citep{Schrijvers1999}.  For a detailed
derivation and extensive discussion of how spectral line profiles change during
p- or g-mode pulsation cycles, we refer to \citet[][Chapter 6]{Aerts2010}.

As appropriate test cases, we consider a few well-studied and bright slowly
rotating B stars, either with coherent pulsations excited by a heat mechanism or
with chemical peculiarity (Bp stars), thus excluding Be stars and evolved B
stars with mass loss.  Chemically peculiar Bp stars and pulsating B stars
co-exist with constant B stars during the main sequence phase
\citep[e.g.][]{Briquet2007}. While the distinction between pulsations and
rotational modulation as cause of the variability of slowly rotating stars is
difficult to make from light curves \cite[e.g.][]{Degroote2011}, it is easy to
do so from high-resolution, high signal-to-noise spectroscopic time series that
we will use here. The properties and the considered data sets of the stars under
study are summarised in Table\,\ref{table1} and discussed in Sect.\,2.1 to 2.3.

\subsection{Spotted B-type stars}

HD\,131120 and HD\,105382 were studied extensively by \citet{Briquet2004}, by
means of high-precision time-series spectroscopy from which the cause of the
variability could be unambiguously determined as rotational modulation.  Doppler
Imaging led to detailed abundance surface maps for Si and He, and to the stellar
parameters listed in Table\,4 of \citet{Briquet2004}, including the rotation
frequency, effective temperature, and gravity listed in
Table\,\ref{table1}. Both stars show two strong and one weak Si spot at their
surface and are moderate rotators with $v\sin i$=60 and 70\,km\,s$^{-1}$ for
HD\,131120 and HD\,105382, respectively. In an attempt to find an explanation
for the surface spots, a magnetic field was sought for both stars.  For
HD\,105382, such a magnetic field has indeed been detected, both in low- and in
high-resolution spectro-polarimetry \citep{Hubrig2006,Alecian2011}.  We are not
aware of any magnetic field detection for HD\,131120, despite efforts to find
one \citep{BriquetAN2007}.

\subsection{$\beta\,$Cep stars}

We considered three well-known $\beta\,$Cep stars in this study. These are stars
that pulsate in low-order p- and g-modes excited by an opacity bump in the
partial ionisation layers of iron-like elements that occurs in their envelope at
a temperature near 2$\times 10^5$\,K \citep{Aerts2010}. Their pulsational
velocity vector's radial component is dominant over the tangential one.  The
first $\beta\,$Cep star considered here is HD\,16582, which is multiperiodic but
has a clear dominant radial mode with moderate amplitude
\citep{Aerts1992,Aerts2006}. The second case is the monoperiodic large-amplitude
radial pulsator HD\,46328 \citep{Saesen2006} and the third class member is the
low-amplitude multiperiodic non-radial pulsator HD\,111123
\citep{Aerts1998,Cuypers2002}. The dominant pulsation frequency and $v\sin i$ of
these stars are listed in Table\,\ref{table1} while their full pulsation
characteristics are available in the above mentioned papers. While HD\,16582 and
HD\,111123 do not possess a detectable magnetic field \citep{Aerts2014},
HD\,46328 does \citep{Hubrig2009}. None of the three stars shows evidence of
detectable rotational modulation in their line profiles.

\subsection{Slowly Pulsating B stars}

The stars HD\,74195 and HD\,181558 are two typical cases of multiperiodic g-mode
pulsators, excited by the same opacity mechanism as the $\beta\,$Cep
stars and better known as slowly pulsating B stars (SPBs hereafter) 
\citep{Aerts2010}. The tangential pulsational
velocity component of such stars is much larger than the radial
component. Their pulsational properties were extensively studied from
multicolour photometry and high-precision spectroscopy by
\citet{DeCatAerts2002}, from which we retrieved the data 
used for the current study. The main difference between these
two stars is that HD\,181558 has one dominant g-mode while HD\,74195 has several
g-modes of similar amplitude. Also, while HD\,181558 has no magnetic field
detected \citep{Aerts2014}, a weak field seems to be present in
HD\,74195 \citep{Bagnulo2012,Hubrig2013}, although it is not accompanied by
a detectable signal of rotational modulation.

\begin{table*}
\caption{Stellar parameters and line-profile properties of seven selected
  variables. The Spectral Types were taken from the Simbad database.}
\label{table1}
\tabcolsep=2pt
\centering                                     
\begin{tabular}{cccccccc}          
\hline\hline\\                     
Star $\rightarrow$ & HD\,131120 & HD\,105382 & HD\,46328 & HD\,16582 &
HD\,111123 & HD\,74195 & HD\,181558 \\
Parameter $\downarrow$ &&&&&&&\\
\hline\\
$T_{\rm eff}$ (K) & 18250 & 17400 & 27500 & 23000 & 27000 & 16000 &
14700\\[0.1cm] 
$\log\,g$ (cgs) & 4.1 & 4.2 & 3.8 & 3.8 & 3.7 & 3.9 & 4.2\\[0.1cm]
Spectral & B7IIIp & B6III & B0.7IV & B2IV & B0.5IV & B3IV & B5III \\
Type & & & & & & & \\[0.1cm]
Variability& Si spots & Si spots & radial & radial & non-radial & non-radial & 
non-radial\\
Type & & & mode & mode & p modes & g modes & gmodes \\[0.1cm]
Selected & \ion{Si}{ii} 4128\,\AA\ & \ion{Si}{ii} 4128\,\AA\ & 
\ion{Si}{iii} 4552\,\AA\ & \ion{Si}{iii} 4552\,\AA\ & \ion{Si}{iii} 4552\,\AA\ &
\ion{Si}{ii} 4128\,\AA\ & \ion{Si}{ii} 4128\,\AA\ \\ 
Line & & & & & & & \\[0.1cm]
$v\sin i$ \ \tablefootmark{a} 
& 60$\pm$5 \ \tablefootmark{b} 
& 70$\pm$5 \ \tablefootmark{b} 
& 15.5$\pm$1.5 \ \tablefootmark{c} 
& 1$\pm$1 \ \tablefootmark{d} 
& 18$\pm$5 \ \tablefootmark{e} 
& 20$\pm$3\ \tablefootmark{e} 
& 9.5$\pm$2.5\  \tablefootmark{f} \\
(km.s$^{-1}$) &&&&&&&\\[0.1cm]
Dominant & 0.63745 &  0.77214 & 4.77153 & 6.20587 & 5.95867 &
0.35745 &  0.80783 \\
Frequency (d$^{-1}$) &&&&&&&\\[0.1cm]
EW (m\AA) &&&&&&&\\
Range & [84,107] & [80,105] & [199,241] & [136,156] & [252,293] & [74,98] &
[100,110]\\
St.Dev. & 6.63 & 6.47 & 7.37 & 4.85 & 6.27 & 4.38 & 2.88\\[0.1cm]
$<v>$ (km.s$^{-1}$) &&&&&&&\\ 
Range & [-4.70,5.85] &[-6.77,9.02] &[-16.19,18.37] &[-5.16,7.09] &
[-2.98,2.75] &[-5.33,4.54] &[-6.66,7.52] \\
St. Dev. & 2.93 & 3.77 & 11.52 & 5.15 & 1.12 & 2.12 & 4.58\\[0.1cm]
$<v^2>$ (km$^2$.s$^{-2}$) &&&&&&&\\
Range &  [754,987] &[987,1496] & [118,502]&  [113,194]& [552,871]&
[319,579] &  [64,220] \\
St.Dev. & 45.8 & 98.2 & 110.1 & 21.5 & 58.7 & 56.8 & 34.2\\[0.1cm]
$<v^3>$ (km$^3$.s$^{-3}$)&&&&&&&\\
Range &  [-9177,13887] &[-25765,25270] & 
[-13635,15814]&  [-3275,2948]& [-7191,6562]&
[-8674,3636] &  [-1468,4688] \\
St. Dev. & 5740 & 11413 & 7741 & 1861 & 2254 & 2305 & 1539\\[0.1cm]
1st zero FT (km.s$^{-1}$)&&&&&&&\\ 
Range & [52.7,60.7] &[43.4,78.8] &[9.1,17.8] & [6.8,15.6] & 
[15.1,46.1] &[6.1,27.5]  &[6.3,26.0] \\
St. Dev. & 1.60 & 6.16 & 2.04 & 2.29 & 8.72 & 5.40 & 6.13\\[0.1cm]
v$_{\rm macro}$ (km.s$^{-1}$) &&&&&&&\\
Range &[11.1,30.0]& [10.8,64.2]& [14.8,21.9]& [14.8,22.7]&
[29.4,54.3]&  [0.0,22.2] & [1.3,25.9]\\
St. Dev.  & 3.64 & 10.50 & 1.47 & 1.73 & 6.67 & 4.70 & 5.54\\[0.1cm]
\hline                                         
\end{tabular}
\tablefoottext{a}{Value deduced from detailed line-profile fitting relying on 
spot or  pulsation models:} 
\tablefoottext{b}{\citet{Briquet2004};}
\tablefoottext{c}{\citet{Saesen2006};}
\tablefoottext{d}{\citet{Aerts1992};}
\tablefoottext{e}{\citet{BriquetAerts2003};}
\tablefoottext{f}{\citet{DeCat2005}.}
\end{table*}

\begin{figure*}[t!]
\begin{center}
\rotatebox{270}{\resizebox{9cm}{!}{\includegraphics{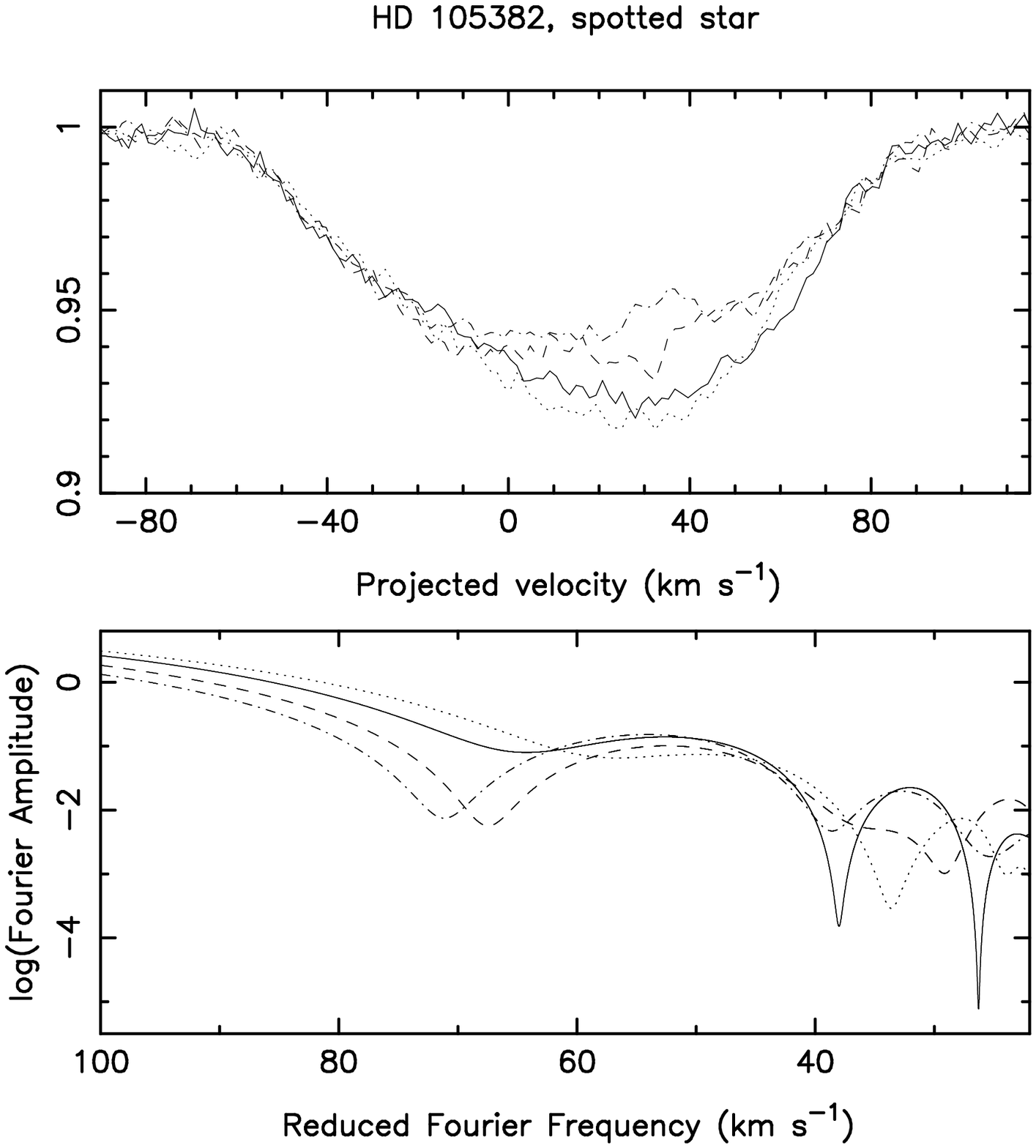}}}\hspace{0.5cm} 
\rotatebox{270}{\resizebox{9cm}{!}{\includegraphics{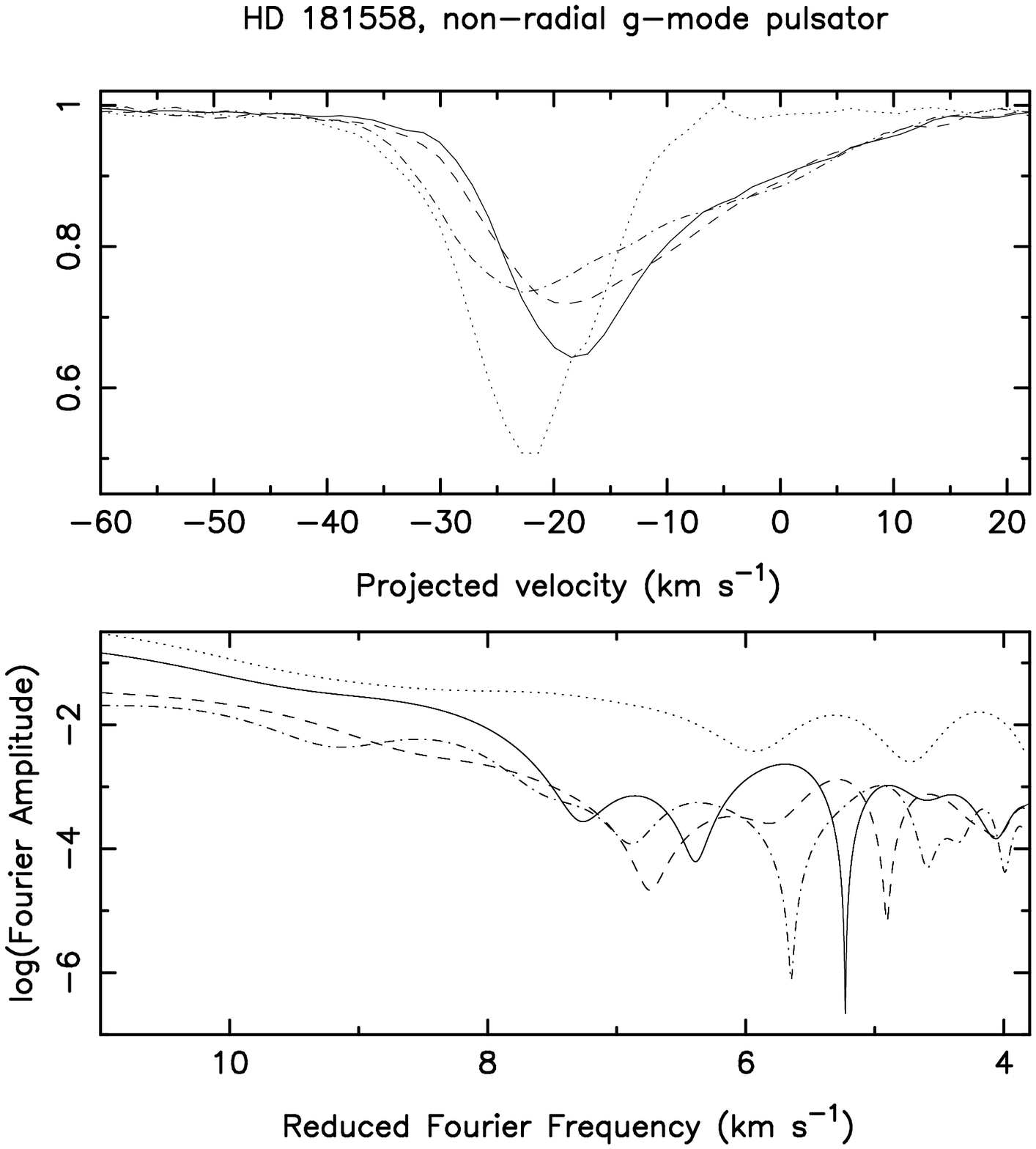}}}\\[0.5cm]
\rotatebox{270}{\resizebox{9cm}{!}{\includegraphics{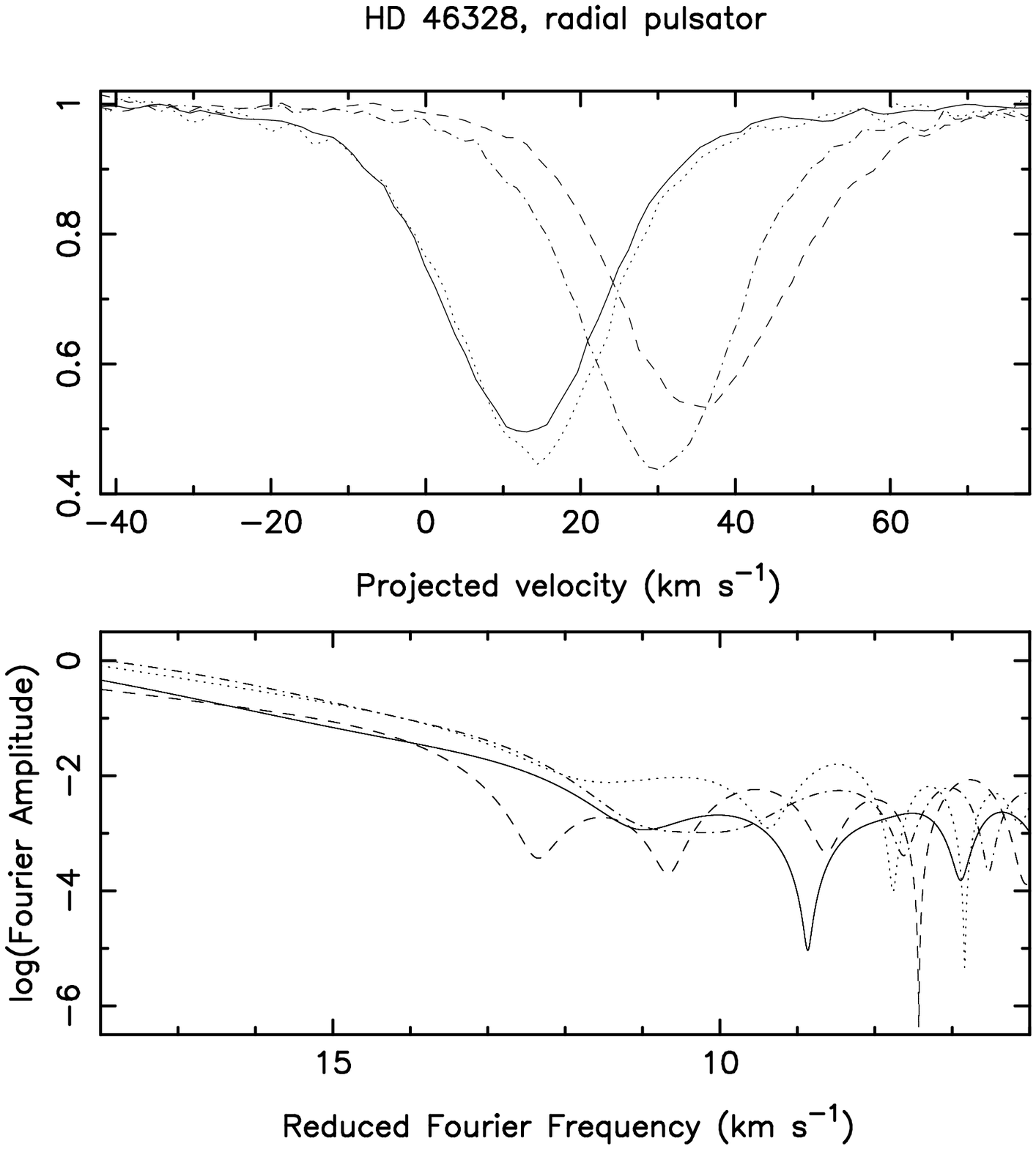}}}\hspace{0.5cm} 
\rotatebox{270}{\resizebox{9cm}{!}{\includegraphics{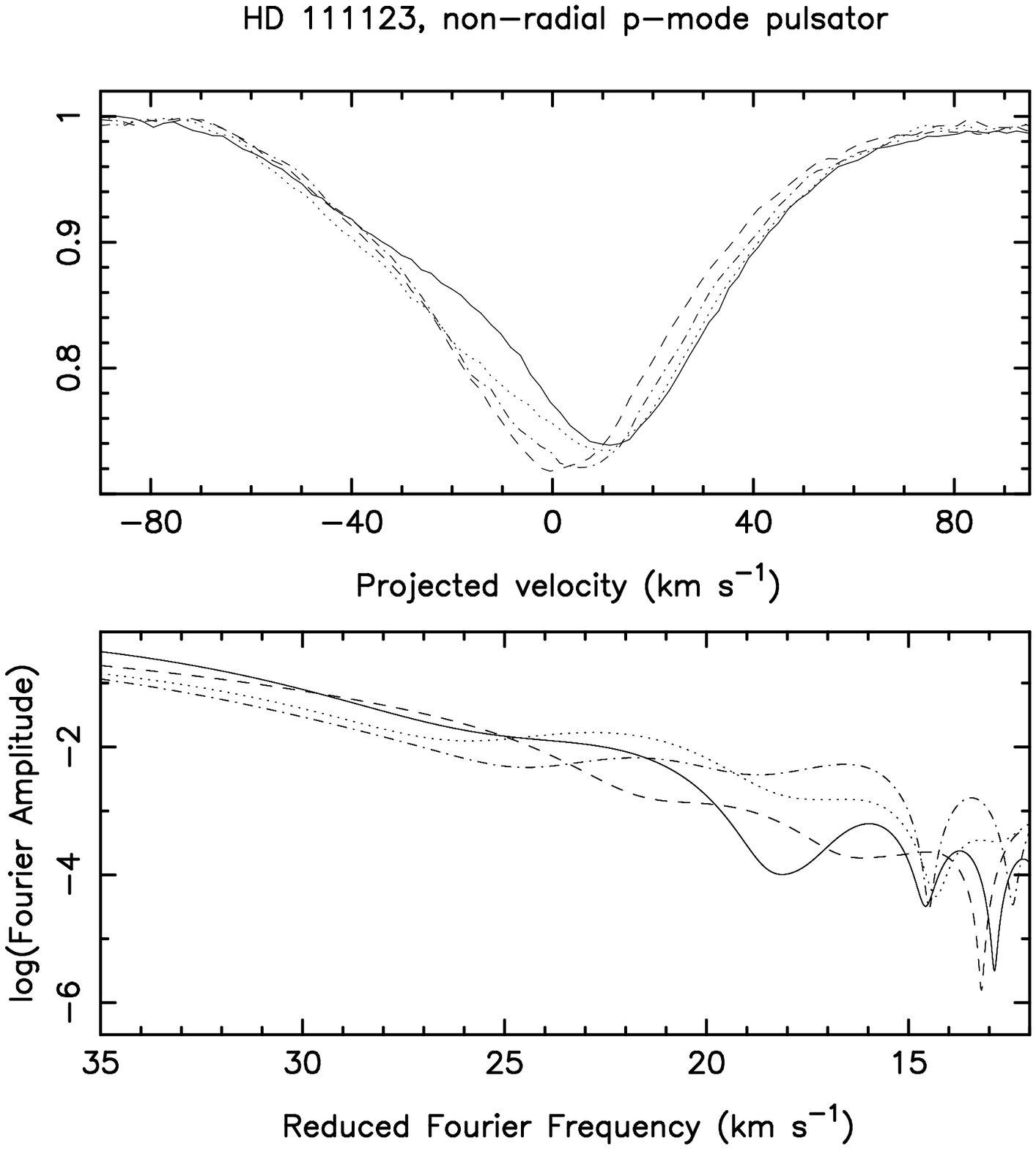}}}
\end{center}
\caption{The Fourier transforms (bottom panels) of four selected profiles (top
  panels) for four of the stars. The abscissa of the bottom panels shows the FT
  frequencies transformed to velocity units following \citet{Jankov1995}. }
\label{FTs}
\end{figure*}

\subsection{The measured line-profile variations}

For the three hottest pulsators studied here, we considered the \ion{Si}{iii}
line at 4553\,\AA, a well-known spectral line optimally suited to interpret the
pulsational variability of $\beta\,$Cep pulsators \citep{AertsDeCat2003}. For
the four cooler B stars, we selected the \ion{Si}{ii} line at 4128\,\AA\ (cf.\
Table\,\ref{table1}).  For four of the stars in Table\,\ref{table1}, the
profiles we have selected to work with have been shown in the literature already
in the above-mentioned papers, so we do not repeat them {all} here and refer
the reader to the orginal data sources.  For three {pulsators whose profiles
  are not yet available in the literature we show the observations in
  Fig.\,\ref{LPVs}, along with the profiles of the spotted star HD\,131120 to
  illustrate the contrast in time series between spotted and pulsating B stars.
Three of the panels of this figure give }
the reader a grasp of how stellar pulsations deform the
individual line profiles for typical members of pulsation classes whose
amplitudes in broad-band photometry are typically between 1 and 35 mmag. It can
be seen, particularly for the g-mode pulsators, that pulsation modes can
strongly affect the line profile shapes, with occurrences of sharp narrow lines
at some occasions during the overall pulsation cycle and broad shallow lines at
other phases. {As illustated in the upper left panel of Fig.\,\ref{LPVs}, the
  spots on the surface of B stars tend to deform the line profiles mainly in
  intensity and much less so in velocity.}

It is noteworthy that the CoRoT and {\it Kepler\/} space missions revealed a
huge diversity of OB-star photometric variability with a great variety in
periodicity and at low levels between 10\,$\mu$mag and 1\,mmag. While
asteroseismology has revealed shortcomings in stellar structure models of B
stars from such data \citep[e.g.][]{Degroote2010}, current time-resolved
spectroscopy of these stars, which have visual magnitudes between 6 and 12,
turned out to be of insufficient resolution and signal-to-noise to capture the
line-profile variability of such low-amplitude pulsations \citep{Degroote2012}.

Appropriate line-profile modelling by means of spots or pulsations requires
extensive computations in terms of either spot models or pulsation theory, or
both. A prerequisite to do so is to identify the character of the spots or the
wavenumbers of the pulsation modes, after frequency analysis from line
diagnostics.  {A powerful method to identify pulsation modes from
  line-profile variations in the case of slow to moderate rotators, is the
  moment method \citep{Aerts1992}. It is based on the idea, inherited from
  statistics, that any distribution function is fully characterised if one knows
  its moments. Hence, a theoretical expression was computed for the moments of a
  line profile based on the eigenvectors and eigenfrequencies deduced from the
  theory of stellar pulsation. The moments vary periodically in time according
  to the pulsation modes of the star. It was shown that the three lowest-order
  moments, denoted as $<v>$, $<v^2>$, and $<v^3>$ and normalised with respect to
  the equivalent width of the line, are sufficient to identify the pulsation
  modes that cause their periodic variability. The theoretical expressions of
  the moments for a monoperiodic pulsator are available in \citet{Aerts1992},
  while those for the generalised version in case of a multiperiodic pulsator
  are derived in \citet{Mathias1994}. The first moment, $<v>$, is a measure of
  the centroid of the line profile and it varies sinusoidally around value zero
  according to the frequencies of the pulsation modes. The second moment,
  $<v^2>$, which is a measure of the width of the line profile, varies with the
  pulsation frequencies and twice their value around a constant value determined
  by the rotational, pulsational, and intrinsic width of the line.  The third
  moment, which measures the skewness of the profile, varies around value zero
  according to once, twice, and three times the pulsational frequencies. Mode
  identification, along with estimation of $v\sin i$ and of the inclination
  angle $i$, is then achieved by minimising the difference between the measured
  moment values and those predicted by theory as a function of the wavenumbers
  $(\ell,m)$ of the modes.  For a more detailed explanation of the moment
  method, we refer to \citet{Zima2008} and \citet[][Chapter
  6]{Aerts2010}.\footnote{The software to perform spectroscopic mode
    identification and achieve line-profile modelling for non-radial pulsators
    is freely available from {\tt
      http://www.ster.kuleuven.be/$\sim$zima/famias/}} } We computed the three
moments for all the line profiles of the seven stars. The range of these moment
values, as well as the standard deviation of these line diagnostics, are
indicated in Table\,\ref{table1}.  {We come back to the temporal variability
  of the moments in Sect.\,\ref{timevariations}.  }
\subsection{Rotational velocities from spot modelling or 
asteroseismology}

The way to estimate the ``true'' $v\sin i$ in the case of rotational modulation
is to model time-resolved profiles covering the rotation period, assuming spots
and/or differential surface rotation. For stars cooler than spectral type B,
current methods are based on the FT coupled to spot models of differentially
rotating stars \citep[e.g.][and references
  therein]{Reiners2002,Reiners2003,Ammler2012}.  For HD\,105382 and HD\,131120,
\citet{Briquet2004} obtained $v\sin i=70\pm 5$ and $60\pm 5$\,km\,s$^{-1}$,
respectively, from modelling of He and Si surface spots assuming constant
surface rotation.  For the case of pulsators, one must perform spectroscopic
mode identification  {as explained in the previous section.}
Only when the time series of line-profile variations is 
modelled properly according to the correct physical processes can one get
an appropriate estimate of the true $v\sin i$ of the star, along with the other
pulsational velocity parameters. 

For the stars in our sample, the $v\sin i$
obtained from spot or pulsation modelling is listed in Table\,\ref{table1}.  As
can be seen, both an under- and overestimation of $v\sin i$ deduced from the 1st
zero of the FT may occur if one ``downgrades'' the level of the line-profile
modelling by considering only the two parameters $v\sin i$ and $v_{\rm macro}$
instead of the complete velocity vector due to all the detected and identified
pulsation modes.

Given that the cases with time-resolved spectroscopy of sufficient time base and
quality to perform the full mode identification are scarce and limited to very
bright stars (visual magnitude typically below 6), it is useful to assess the
performance of a ``downgraded'' analysis based on $v\sin i$ and $v_{\rm macro}$
when the aim is to evaluate stellar evolution models from large samples of
stars. For such applications, line-profile variability is usually ignored and
one tends to work with snapshot spectra only, rather than time-resolved
spectroscopy. As shown in Fig.\,\ref{LPVs} and listed in Table\,\ref{table1},
this paper concerns slow to moderate rotators; for
fast rotators, the pulsational signal in the line profiles gets smeared out due
to the large rotational broadening whenever the latter becomes dominant over the
pulsational broadening.


\section{Results of the Fourier Transform method}

Following the results of SDH14 and keeping in mind that pulsations lead to
radial as well as tangential velocities, we considered a radial-tangential
description of macroturbulence to try and capture the line-profile variability
in such a formalism.  The time-resolved spectroscopy of the seven stars
introduced in the previous section were subjected to the FT method with the aim
to determine $v\sin i$, and subsequently $v_{\rm macro}$, for each of the
selected line profiles in the time series. In total we analysed 866 Si line
profiles as if spots or pulsations are absent 
{and we derived
$v_{\rm macro}$ in this approximation.
}

We first of all used the {\tt iacob-broad} tool as described in SDH14 to perform
the line-broadening analysis. This tool has a fixed linear limb-darkening
coefficient of 0.6 and makes four sets of computations: 
$v\sin i$ from the first zero of the FT, $v\sin
i$ from a GOF where $v_{\rm macro}$ is assumed zero, $v\sin i$ and $v_{\rm
  macro}$ from a GOF assuming both are free parameters, and $v_{\rm macro}$
resulting from a GOF where the $v\sin i$ is fixed to the value corresponding to
the first zero of the FT.  { In the current version of {\tt iacob-broad}, the
radial and tangential contributions to the macroturbulent velocity are assumed
to be equal.}
In the rest of this paper, we consider the $v\sin i$
from the first zero of the FT derived from {\tt iacob-broad} as the baseline
result.

In addition, we used two other implementations of the FT method
by \citet{Jankov1995} and \citet{Piters1996} to assess the robustness of the
$v\sin i$ determination from the first zero of the Fourier transform of the
selected line profile.  All three implementations are based on the same
principle that the rotational broadening of a spectral line can be distinguished
from other additional broadening mechanisms thanks to the property that the
convolution of various broadening functions in the wavelength domain transfers
into a multiplication in the Fourier domain.

Under the assumptions that one deals with uniform surface rotation, a
homogeneous surface temperature and surface chemical composition, and a
spherical star with linear limb darkening, $v\sin i$ can in principle be deduced
from the FT alone, without performing detailed modelling of the overall line
profile shape.  A further assumption is that all occurring broadening mechanisms
act independently of each other.  Under these circumstances, the zeros
of the Fourier transform of the profile are inversely proportional to 
$v\sin i$ \citep[e.g.,][for the proportionality coefficients]{Dravins1990}.
It is then customary to deduce $v\sin i$ from the first zero of the FT.

The caveat we want to highlight in this work is that the occurrence of surface
inhomogeneities or of pulsations introduces line asymmetries, which transform
into saddle points or local minima unrelated to $v\sin i$, in the Fourier
transform of the line profile. This is illustrated for four different profiles
observed during the variability cycle for four of the stars in
Fig.\,\ref{FTs}. It can be seen that the ``first'' zero point not only shifts
from one spectrum to another one, but even more importantly that the difference
between a saddle point, a local flat minimum, and the first `true'' minimum can
be hard to judge. This difficulty lies at the origin of inaccurate $v\sin i$
estimates in the case of line-profile variable stars, as we will illustrate in
the following.

\subsection{Effect of limb darkening on the FT method}

\begin{figure}
\begin{center}
\rotatebox{270}{\resizebox{5cm}{!}{\includegraphics{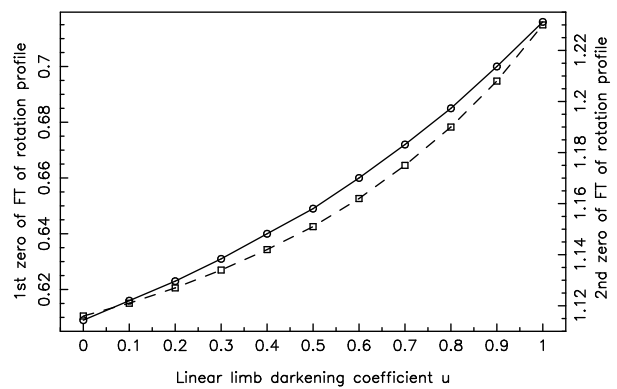}}}
\end{center}
\caption{The first (left $y$-axis, full line) and second (right $y$-axis, dashed
  line) zero points of the amplitude of the FT of a rotation profile as a
  function of the linear limb darkening coefficient. Figure produced from
  Table\,1 in \citet{Jankov1995}.}
\label{jankov}
\end{figure}

As already stressed by \citet{Jankov1995}, the zero points of the FT depend on
the limb darkening of the star. Limb darkening coefficients are available in the
literature as a function of $T_{\rm eff}$, $\log\,g$, and metallicity of the
star, for various wavelength regimes \citep[e.g.,][and references
  therein]{Claret2011}. A typical value adopted for early B-type dwarfs in the
case of a linear law is $u=0.35$
\citep{Slettebak1949,Aerts1992}.  The temperature varies
across the stellar surface and in time due to stellar pulsations. However, as
already mentioned above, the changes of line profiles due to pulsations are
mainly caused by the local velocity variations 
on the stellar surface and much less
by the local temperature and gravity variations \citep{DeRidder2002}. Moreover,
\citet{Aerts1992} investigated the effect of changing the adopted limb darkening
when interpreting line-profile variations of hot massive stars and found that
the approximation of a constant linear limb-darkening law during the entire
pulsation cycle is largely sufficient to interpret the time-series spectroscopy
adequately.  We thus work in this linear limb darkening approximation in the
present paper, but we do show the effect of choosing a different value for the
limb darkening coefficient.

In the recent literature, the limb darkening dependence in the derivation of
$v\sin i$ has often been ignored in the sense that the limb darkening was either
not taken into account \citep{Piters1996} or fixed at the ``standard'' value of
$u=0.6$ as adopted by \citet{Gray2005}. However, this standard value is only
appropriate for cool stars and less so for hot stars as we study here. The
effect of limb darkening on the derived value of $v\sin i$ is not negligible but
it is known.  For convenience, we reproduce the table with the first and second
zero points of the amplitude of the FT as given by \citet{Jankov1995} in
Fig.\,\ref{jankov}. It can be seen from this figure that the factor between the
first zero of the FT for a fully darkened disk and for a constant intensity
across the stellar surface amounts to 0.85. This figure, or Table\,1 in
\citet{Jankov1995}, can be used to rescale the stated $v\sin i$-values in the
literature that have been derived without limb darkening as in
\citet{Piters1996,Groot1996} or with a standard value of 0.6 for the linear
coefficient as in SDH14, while $u$ should be adapted to the temperature, gravity
and metallicity of the star, as well as to the laboratory wavelength considered.
Of these four, the temperature and wavelength are the two most important ones to
influence $u$.  In early applications (cf.\ \citet{Jankov1995} for a
discussion), the derivation of $v\sin i$ and $u$ from the FT was done
simultaneously from the first two zeros.  Indeed, the ratio between the first
and second zero point as shown in Fig.\,\ref{jankov} can be used to derive the
most appropriate value of $u$.  In modern applications, however, one works the
other way around: by fixing the most appropriate value for $u$ according to the
temperature of the star and the wavelength of the spectral line, one can assess
whether the surface rotation is differential or not \citep[e.g.][for numerous
  applications to cool stars]{Reiners2003,Ammler2012}.

Given that the linear limb darkening coefficient was fixed to $u=0.6$ in SDH14
and that we want to quantify the effects of stellar pulsations or spots on the
derived $v\sin i$ and $v_{\rm macro}$, we adopted the 
result from {\tt iacob-broad} relying
on $u=0.6$ for the comparisons between the different implementations of the
FT method, but we do show below the effect of using a different value of $u$ as
well. 

\subsection{The Fourier-Bessel method}

Rather than considering a pure Fourier transform, \citet{Piters1996} used the
Fourier-Bessel transform (FBT) of the line profile. The idea behind this is that
the Fourier transform of a pure rotational profile, which has an ellipsoidal
shape in the absence of limb darkening, is proportional to a first-order Bessel
function and that the Bessel transform of this Fourier transform has a maximum
at $v\sin i$. The method by \citet{Piters1996} is implemented with the aim to
indicate this maximum by the user, from a graphical user interface. The code has
the option to be applied to a large number of spectra, as was done and
illustrated in \citet{Groot1996} for a sample of F-type stars.

In order to compare the results of the FBT with those obtained from the other
implementations, we must scale the velocity value derived from the maximum
of the FBT, in order to take into account the effect of limb darkening, which we
did according to the factor 1.084 as can be deduced from Fig.\,\ref{jankov}.

\begin{figure*}[t!]
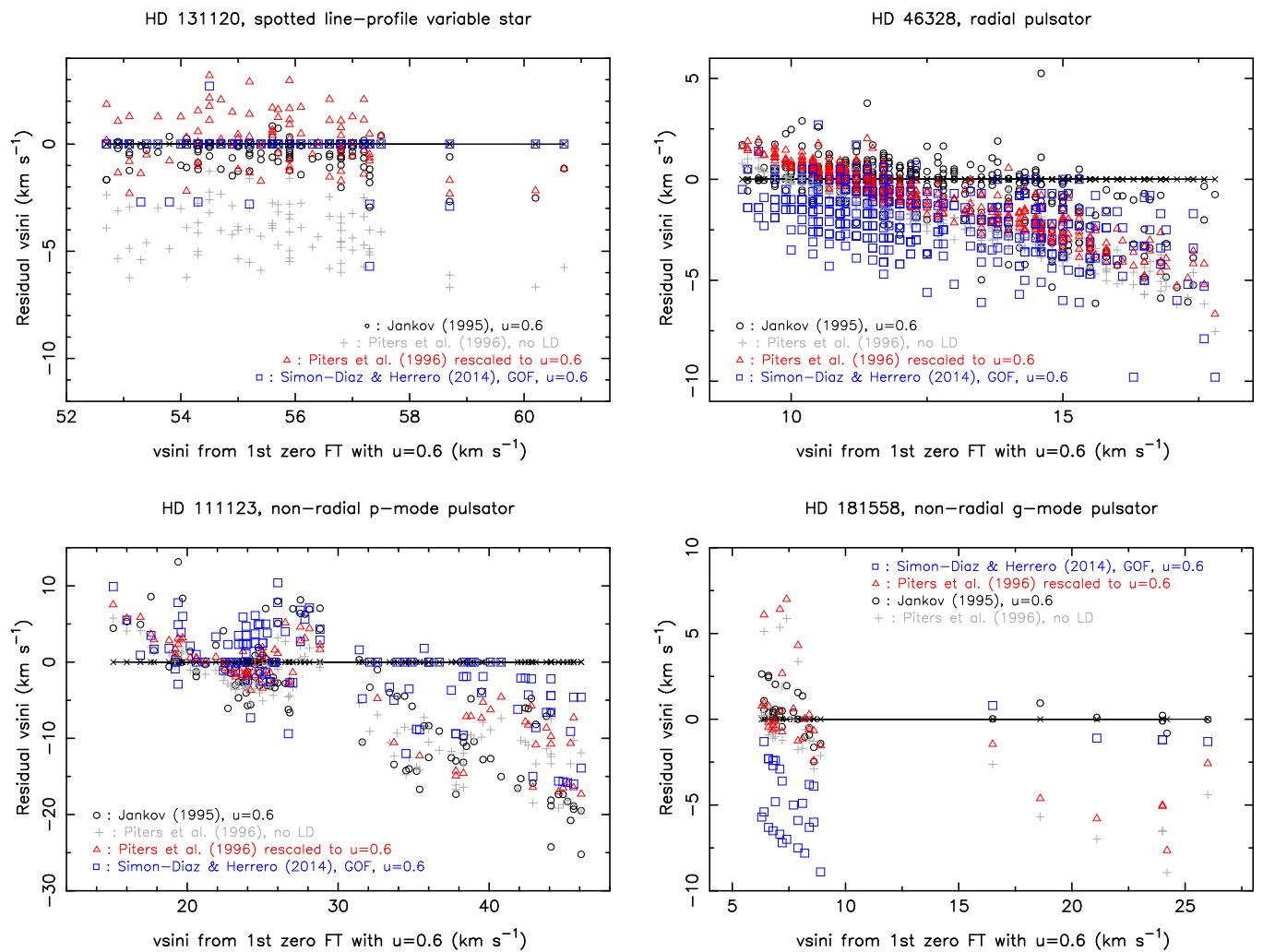

\begin{center}
\rotatebox{270}{\resizebox{6.5cm}{!}{\includegraphics{hd131120_FTFBT.eps}}}\hspace{0.5cm} 
\rotatebox{270}{\resizebox{6.5cm}{!}{\includegraphics{hd46328_FTFBT.eps}}}\\[0.5cm] 
\rotatebox{270}{\resizebox{6.5cm}{!}{\includegraphics{hd111123_FTFBT.eps}}}\hspace{0.5cm} 
\rotatebox{270}{\resizebox{6.5cm}{!}{\includegraphics{hd181558_FTFBT.eps}}}
\end{center}
\caption{Comparison between the $v\sin i$ derived from the FT method with
  $u=0.6$ (full line) and from the FBT and GOF methods as indicated. Colour
  figure only available in the online version of the paper.}
\label{FTFBT}
\end{figure*}

\subsection{Comparison of the results for $v\sin i$}

The authors used three implementations of the F(B)T method to derive $v\sin i$
for all the profiles, independently from each other. This is important to
realise, because the results require manual intervention in the sense that the
user must select the position of the first extremum. As already said, this is
not trivial for the FT method in the cases where saddle points and local minima
occur. The {\tt iacob-broad} and \citet{Jankov1995} tools are different
implementations of the same method and were applied by two of the authors
independently (SSD and CA) to take into account the subjectivity of selecting
the 1st minimum; the \citet{Piters1996} implementation is less subjective in the
choice of the maximum of the FBT because it is less dependent on the
line-profile variability, but the exact position of the maximum is less clear
and one needs to rescale the result to take into account the limb darkening
effect. 

We show the comparative results for 
{ four} of the stars in
Fig.\,\ref{FTFBT}.  We adopted the results of FT using the {\tt iacob-broad}
tool designed by SDH14, where $u=0.6$ was fixed, as the baseline to compare with
(crosses connected by full line in the panels).  In addition to the results from
the FT method, we also show the results obtained from a GOF approach to
determine $v\sin i$ and $v_{\rm macro}$ from {\tt iacob-broad}, as explained in
much detail in SDH14.  The {upper left panel of Fig.\,\ref{FTFBT}} 
shows the results
for the spotted star HD\,131120. It can be seen that the FT versus GOF results
for $v\sin i$ from {\tt iacob-broad} (full line versus squares) are in almost
perfect agreement except for a few profiles in the time series. There is good
agreement with the results from the \citet{Jankov1995} method and also with the
results from the FBT method once these are scaled to $u=0.6$ according to
Fig.\,\ref{jankov}. If one keeps in mind that the $v\sin i$ estimate is expected
to reach a precision of a few \%, then the comparative results for HD\,131120
are excellent, {keeping in mind that spot modelling led to $v\sin i=60\pm
5$\,km\,s$^{-1}$. }

{For the large-amplitude radial pulsator HD\,46328, the $v\sin i$ values
  derived from the different codes have quite good agreement, although we find
  the GOF results to deviate considerably more from the F(B)T results in
  comparison with the case of a spotted star. This is understandable from the
  fact that a radial pulsation mode dominantly acts on the central part of the
  line profile where it introduces broadening and slight skewness due to the
  radial pulsation velocity, while leaving the outer line wings almost
  unaltered. The effect of such profile deformation remains limited in the
  Fourier transform, but the GOF tries to enforce a good fit to the
  pulsationally broadened profiles. { Deviations from the 
$v\sin i=15.5\pm
  1.5$\,km\,s$^{-1}$ that was obtained from pulsational modelling of the
  line profiles are quite large for all the FT methods.}

The situation is worse for the non-radial multiperiodic p-mode
pulsator HD\,111123.  This is because the non-radial modes lead to broad 
asymmetric line
wings and hence saddle points occur in the F(B)T, which may imply ambiguity or
even an incorrect choice of the position of the first zero point of the FT and
uncertainty in the position of the maximum of the FBT.  In addition to that, the
results from the FT and GOF from {\tt iacob-broad} differ appreciably, as can be
deduced from the spread of the open squares with respect to the full line in the
middle panel of Fig.\,\ref{FTFBT}. { Some estimates are more than a factor
  two different from the $v\sin i=18\pm 5$\,km\,s$^{-1}$ deduced from
  appropriate pulsational modelling.} 

Finally, the results for HD\,181558 show yet another pattern, with a clump for
the F(B)T methods around a value of $v\sin i \sim 8$\,km\,s$^{-1}$, except for a
few outliers at much higher $v\sin i$ between 15 and 25\,km\,s$^{-1}$ {compared
to the $v\sin i=9.5\pm 2.5$\,km\,s$^{-1}$ deduced from pulsational modelling.}
  These
points correspond to the few profiles in Fig.\,\ref{LPVs}, which are clearly
broader than the other ones due to particular beating of the non-radial modes.

Note that, although the FBT method was not designed to provide modelling of the
other broadening mechanisms occuring in addition to rotation, particularly not
when anisotropic macroturbulence is present \citep{Piters1996}, it does not
perform worse in the estimation of the correct $v\sin i$ than the FT method,
which is supposed to have good capacities in such situations \citep{Gray2005}.

\begin{figure*}
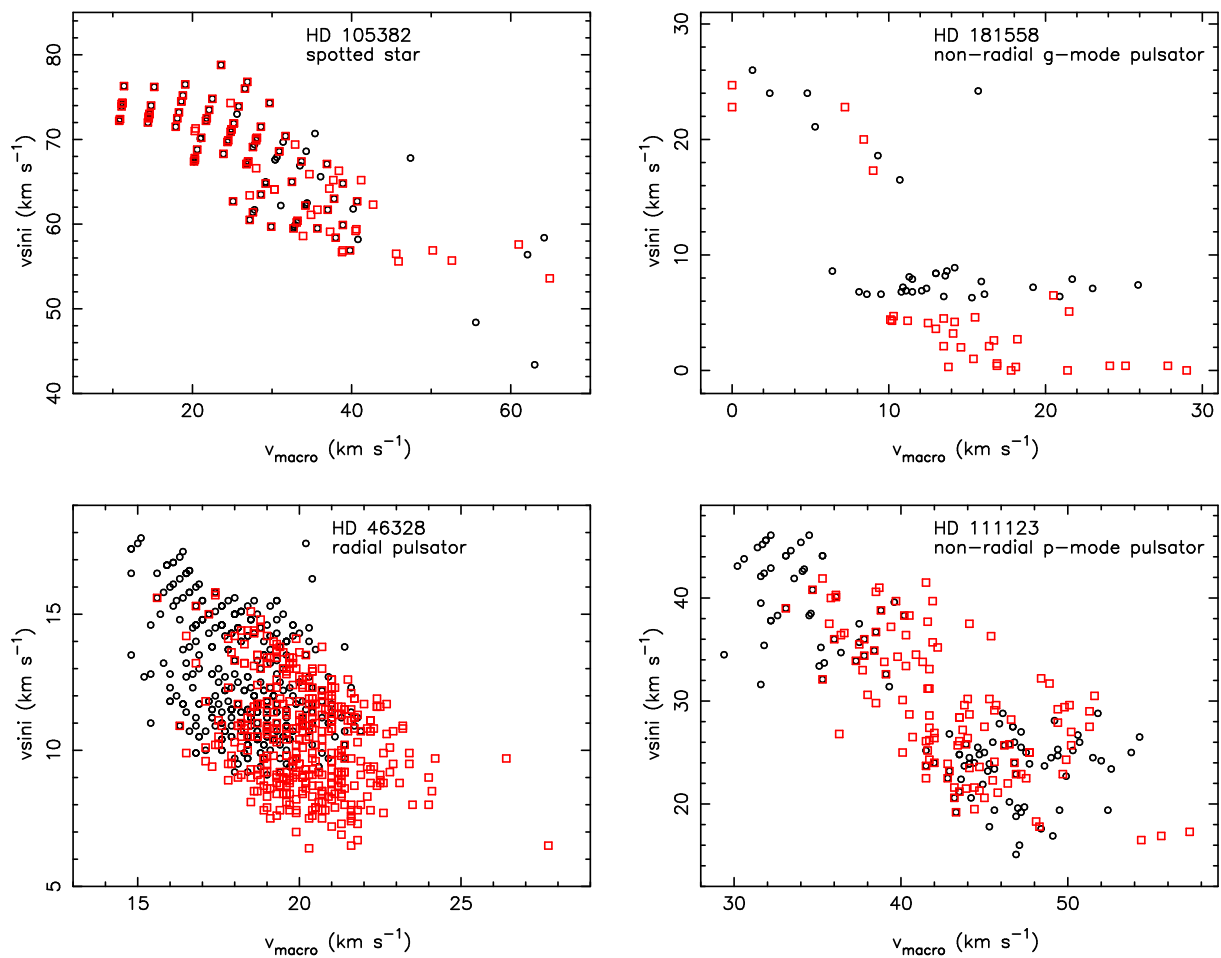

\begin{center}
\rotatebox{270}{\resizebox{6cm}{!}{\includegraphics{hd105382_FTvmacro.eps}}}\hspace{0.5cm} 
\rotatebox{270}{\resizebox{6cm}{!}{\includegraphics{hd181558_FTvmacro.eps}}}\\[0.5cm]
\rotatebox{270}{\resizebox{6cm}{!}{\includegraphics{hd46328_FTvmacro.eps}}}\hspace{0.5cm} 
\rotatebox{270}{\resizebox{6cm}{!}{\includegraphics{hd111123_FTvmacro.eps}}}
\end{center}
\caption{The projected rotation velocity as a function of the deduced
macroturbulent
  velocity. Black circles: $v\sin i$ deduced from the FT method and fixed for
  subsequent derivation of $v_{\rm macro}$ through GOF; red squares: both values
  deduced from a simultaneous two-parameter GOF approach. Colour
  figure only available in the online version of the paper.}
\label{FTvmacro}
\end{figure*}


\section{Temporal behaviour of $v\sin i$ and $v_{\rm macro}$\label{timevariations}} 

As can be deduced from Fig.\,\ref{FTFBT}, the range of $v\sin i$ across the
rotation phase of HD\,131120 is only 8\,km\,s$^{-1}$, while it reaches about
35\,km\,s$^{-1}$ during the pulsation cycle of the dominant mode for HD\,111123.
We computed the temporal range and standard deviation of $v\sin i\,(t)$ and of
$v_{\rm macro}\,(t)$ for the seven stars and included them in
Table\,\ref{table1}.  A general remark is that we did not treat the
microturbulence separately in this work. In this sence, what we list as range of
$v_{\rm macro}\,(t)$ in Table\,\ref{table1}, is an upper limit for
the total amount of { turbulent line-broadening}.

First of all, we come to the conclusion that the $v\sin i\,(t)$ range over the
rotation or pulsation cycle can either be very narrow (HD\,131120) or broad
(HD\,111123) depending on the nature of the spots or pulsations.  We cannot
provide a general recipe of how $v\sin i\,(t)$ behaves during the variability
cycle from this small sample of stars. Clearly, the range in $v\sin i\,(t)$ not
only depends on the type of variability but is quite different within one class
(cf.\ HD\,131120 versus HD\,105382).

The range in $v\sin i\,(t)$ values obtained during the rotation or pulsation
cycle has repercussions on the values derived for $v_{\rm macro}\,(t)$ as
well. We followed the approach of SDH14 to derive the $v_{\rm macro}\,(t)$ from
either a GOF approach after fixing the $v\sin i\,(t)$ deduced from the FT
method, or from a GOF for the two quantities simultaneously. The results are
illustrated for four of the seven stars in Fig.\,\ref{FTvmacro}, where the
circles are the results for the FT+GOF method and the squares for the GOF method
to fit both $v\sin i\,(t)$ and $v_{\rm macro}\,(t)$.  It can be seen that the
two methods agree very well for the spotted star HD\,105382, which had a broad
range of $v\sin i\,(t)$ values resulting from the FT method. For the three
pulsators, we find a lower $v\sin i\,(t)$ value and an accompanying higher
$v_{\rm macro}\,(t)$ value for the GOF method compared to the FT method. This is
logical if one keeps in mind that pulsational broadening during the cycle
affects the blue and red line wings in an asymmetrical way
\citep[cf.\ Fig.\,\ref{LPVs} and][Chapter 6]{Aerts2010} and the GOF tries to
find the best compromise for the red and blue wing, while spots typically give
rise to more local variability in the line, which implies less asymmetry in the
line wings 
{ (cf.\ Fig.\,\ref{LPVs}).  It was already stressed in SDH14 and in
  \citet{Sundqvist2013}, as well as in earlier work, that the GOF approach to
  fit $v\sin i$ and $v_{\rm macro}$ simultaneously may lead to a large
  degeneracy in the parameter space, particularly for slow rotators.  This
  becomes obvious also in our study of the seven stars from Table\,\ref{errors},
  where we list the maximal uncertainty for $v\sin i\,(t)$ and for $v_{\rm
    macro}\,(t)$ when determined from a GOF analysis, as well as the uncertainty
  for these two quantities averaged over the total time series.
}

\begin{table}
\caption{Maximal error and error averaged over the time series spectroscopy 
encountered for the $v\sin i$ and $v_{\rm macro}$ estimates when
  using the GOF approach to determine both quantities.}
\label{errors}
\tabcolsep=2pt
\centering 
\tabcolsep=5pt                                    
\begin{tabular}{|l|r|r|r|r|}          
\hline\\[-10pt]                  
 Star & 
\multicolumn{2}{c|}{Error $v\sin i$ (km\,s$^{-1}$)} & 
\multicolumn{2}{c|}{Error $v_{\rm macro}$ (km\,s$^{-1}$)}\\
\hline\\[-10pt]
 & Maximum & Average& Maximum & Average \\
\hline\\[-10pt]
HD\,131120 & 15.8 & 3.9 & 29.0 & 12.0 \\ 
HD\,105382 & 57.6 & 7.6 & 61.0 & 17.3 \\
HD\,46328 & 12.2 & 2.3 & 13.9 & 2.0 \\
HD\,16582 & 12.7 & 1.9 & 10.8 & 1.8 \\
HD\,111123 & 6.6 & 2.9 & 9.2 & 3.0 \\
HD\,74195 & 18.9 & 2.8 & 42.4 & 4.9 \\
HD\,181558 & 5.6 & 3.0 & 6.3 & 1.6 \\
\hline
\end{tabular}
\end{table}

We note here in passing that the
implementation of \citet{Piters1996} allows the user to ``symmetrize'' the line
profile, by mirroring either the left or the right part with respect to the line
centre. We did not apply this tool in this work, because it is an
inappropriate data manipulation in the case where one {\it knows\/} that line
asymmetries occur due to pulsations and/or spots, rather than being due to line
blending in one part of the line for which this manipulation is well justified.

\begin{figure}
\begin{center}
\rotatebox{270}{\resizebox{9.2cm}{!}{\includegraphics{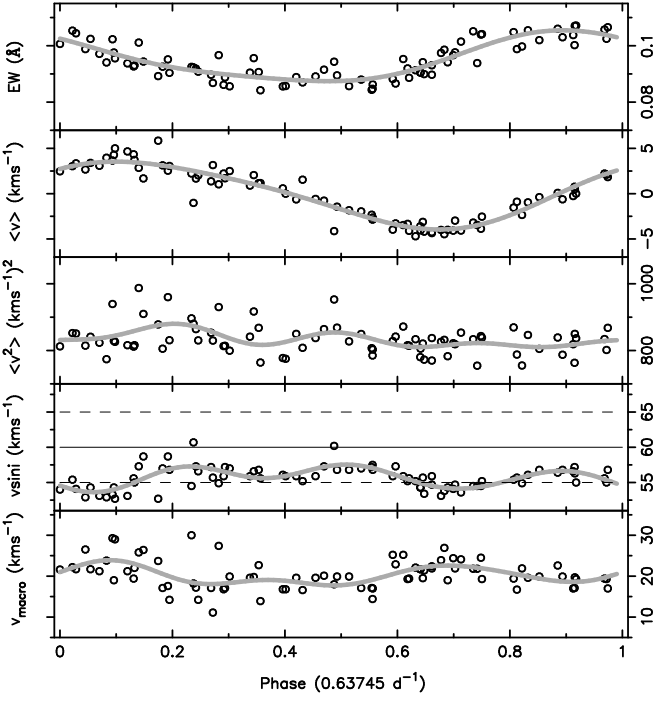}}}\\[0.5cm]
\rotatebox{270}{\resizebox{9.2cm}{!}{\includegraphics{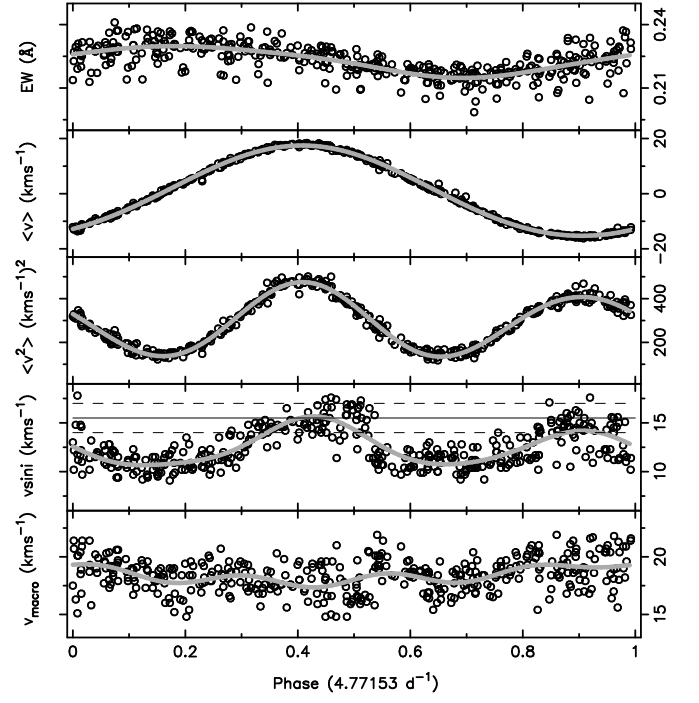}}}
\end{center}
\caption{Line diagnostics of two rotationally modulated variables due to Si
  spots. The horizontal lines indicate the true $v\sin i$ from spot modelling
  (full) and its 1$\sigma$ undertainty (dashed lines).  Upper: HD\,131120,
  Lower: HD\,105382.}
\label{vsinivmacro.spots}
\end{figure}

\begin{figure}
\begin{center}
\rotatebox{270}{\resizebox{9.2cm}{!}{\includegraphics{hd46328_vsinivmacro.eps}}}\\[0.5cm]
\rotatebox{270}{\resizebox{9.2cm}{!}{\includegraphics{hd111123_vsinivmacro.eps}}}
\end{center}
\caption{Line diagnostics of two p-mode pulsators. Upper: HD\,46328, which is a
  monoperiodic radial pulsators, Lower: HD\,111123, which is a multiperiodic
  non-radial pulsator. The horizontal lines indicate the true $v\sin i$ from
  modelling of the line-profile variations in terms of pulsation theory (full)
  and its 1$\sigma$ undertainty (dashed lines). }
\label{vsinivmacro.pmodes}
\end{figure}

\begin{figure}
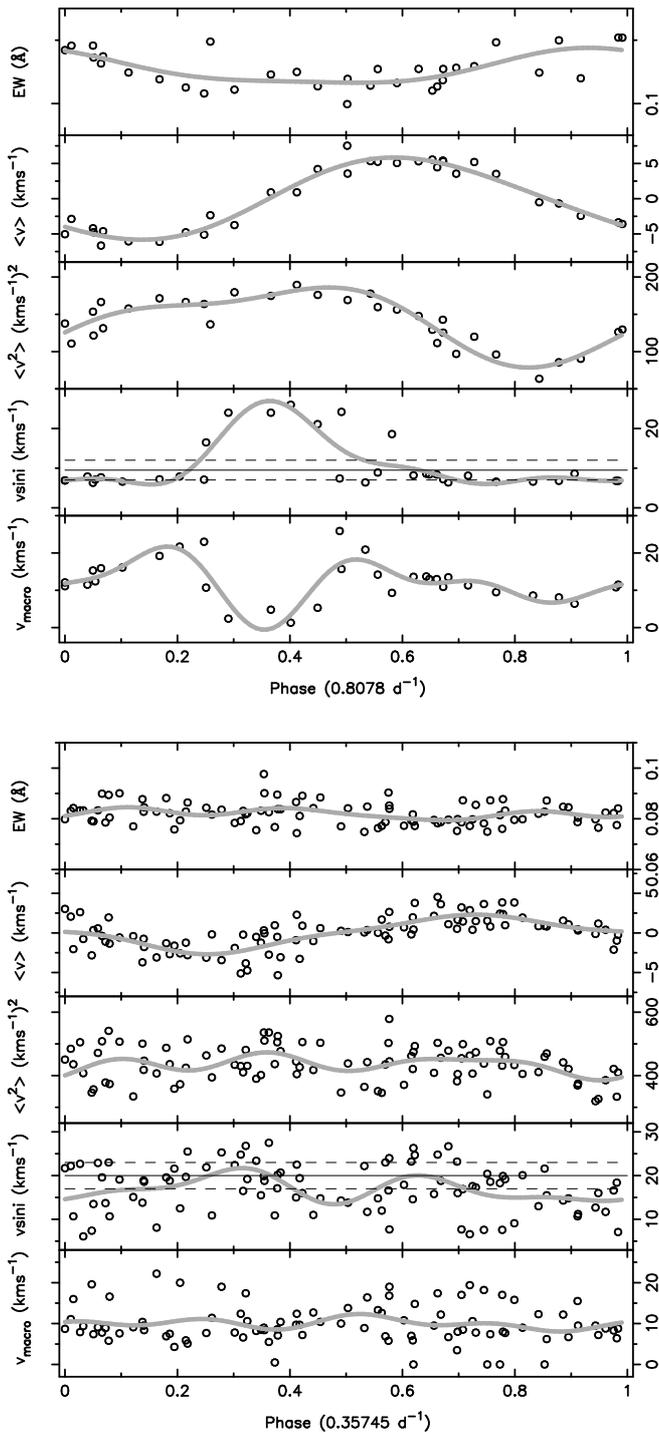

\begin{center}
\rotatebox{270}{\resizebox{9.2cm}{!}{\includegraphics{hd181558_vsinivmacro.eps}}}\\[0.5cm]
\rotatebox{270}{\resizebox{9.2cm}{!}{\includegraphics{hd74195_vsinivmacro.eps}}}
\end{center}
\caption{Line diagnostics of two g-mode pulsators. Upper: HD\,181558, which has
  a dominant g mode, Lower: HD\,74195, which has various g modes of almost equal
  amplitude. The horizontal lines indicate the true $v\sin i$ from
  modelling of the line-profile variations in terms of pulsation theory (full)
  and its 1$\sigma$ undertainty (dashed lines).}
\label{vsinivmacro.gmodes}
\end{figure}

Subsequently, we performed a frequency analysis of the line moments, as well as
of the time series for $v\sin i\,(t)$ derived { from the FT} and for $v_{\rm
  macro}\,(t)$ { derived from the GOF}.  The results are shown in
Figs\,\ref{vsinivmacro.spots}, \ref{vsinivmacro.pmodes}, and
\ref{vsinivmacro.gmodes}, where we also indicated the ``true'' value of $v\sin
i$ as deduced from detailed line-profile modelling based on spot models or
pulsation theory as given in Table\,\ref{table1} (horizontal lines).

For the two spotted stars, we find the dominant frequency to occur at a value of
three times the rotation frequency for both $v\sin i\,(t)$ and $v_{\rm
  macro}\,(t)$. This is as expected, since these two stars each have three Si
spots on their surface. 
{ As discussed by \citet{Briquet2004}, the moment variations
due to the spots of HD\,131120 and of HD\,105382 
are readily different from those expected for pulsations. Here,}
we find a correlation between the phase behaviour of
$v\sin i\,(t)$ and of the second moment $<v^2>\,(t)$, while $v_{\rm macro}\,(t)$
shows an anti-phase correlation with $<v^2>\,(t)$ and $v\sin i\,(t)$. One needs
a value of $v_{\rm macro}\,(t)$ that is smaller or of the same order than the
rotational broadening to mimic the effects of the spots on the line profile
shape.  That is a reflection of the fact that chemical or temperature
inhomogeneities are not accompanied by velocity signatures but rather change the
emergent local intensity such that the line profiles remain dominantly broadened
by rotation.

{ For all the pulsators, the three moments $<v>$, $<v^2>$, and 
$<v^3>$ reveal
  the pulsation frequencies as already known from the previous studies of these
  stars.  A clear result of the present study is that we recover one of the
  dominant mode frequencies in the time series of $v\sin i\,(t)$ and $v_{\rm
    macro}\,(t)$.}
 Hence, the
pulsational velocity field due to pulsations implies periodicity in the
broadening parameters $v\sin i\,(t)$ and $v_{\rm macro}\,(t)$ when one uses only
those two quantities to mimic the line-profile shapes induced by the pulsations.
Strong correlations between the moments and the $v\sin i\,(t)$ are found for the
radial pulsators HD\,16582 (not shown) and HD46328 (upper panel of
Fig.\,\ref{vsinivmacro.pmodes}), while the periodicity is less clear for these
stars' $v_{\rm macro}\,(t)$. For the non-radial pulsators (lower panel of
Fig.\,\ref{vsinivmacro.pmodes} and both panels of
Fig.\,\ref{vsinivmacro.gmodes}), we find again a clear correlation between the
phase behaviour of $<v^2>\,(t)$ and $v\sin i\,(t)$, while $v_{\rm macro}\,(t)$
has an anti-phase correlation. This reflects that the large pulsational
broadening contribution in $<v^2>\,(t)$ is compensated as much as possible by
adapting $v\sin i\,(t)$ to it, but this leads to too broad line wings given the
ellipsoidal nature of the rotation profile, which is then compensated by a
smaller $v_{\rm macro}\,(t)$ to represent best the shape of the pulsationally
broadened line wings. At the phases where the pulsational broadening leads to
lower $<v^2>\,(t)$, i.e., smaller line width, the $v_{\rm macro}\,(t)$ increases
in order to mimic the broader line wings (cf.\ Fig.\,\ref{LPVs}).
As can be seen in Figs\,\ref{vsinivmacro.pmodes} and
Fig.\,\ref{vsinivmacro.gmodes}), both an under- and overestimation of $v\sin
i\,(t)$ deduced from the 1st zero of the FT may occur if one ``downgrades'' the
level of the line-profile modelling by considering only the two parameters
$v\sin i\,(t)$ and $v_{\rm macro}\,(t)$ instead of the complete velocity vector
due to all the detected and identified pulsation modes.

In conclusion, for both the p- and g-mode pulsators studied here, the
anti-correlation between $v\sin i\,(t)$ and $v_{\rm macro}\,(t)$ caused by the
pulsations is clear. The {\it macroturbulent velocities needed to mimic the
  pulsational broadening are of the same order or somewhat larger than the true
  rotational broadening}.

\begin{figure}[t!]
\begin{center}
\rotatebox{270}{\resizebox{6cm}{!}{\includegraphics{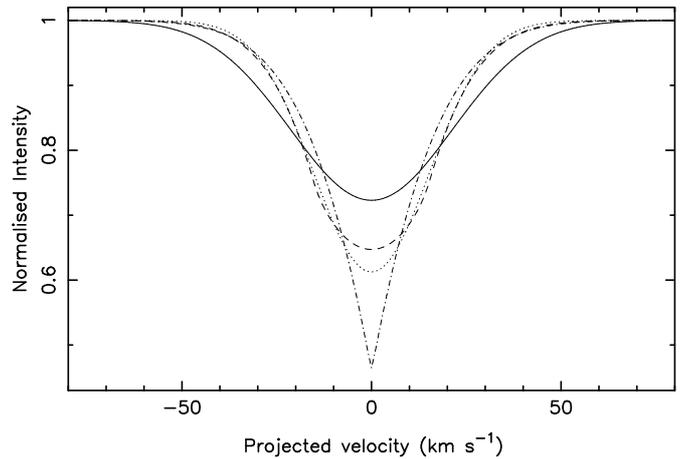}}}
\end{center}
\caption{ { Simulated line profiles of equal equivalent width for $v\sin
    i=15\,$km\,s$^{-1}$ and a macroturbulent velocity of 20\,km\,s$^{-1}$ for an
    isotropic (full line), a pure radial (dashed line), a pure tangential
    (dashed-dot line), and a radial-tangential with equal contributions of the
    two (dotted line).}}
\label{isoRT}
\end{figure}

\section{Conclusions}
\label{conclusions}

Various high-precision time-resolved spectroscopic studies performed the past
two decades have revealed that a significant fraction of the slowly to
moderately rotating 
{ ($v\sin i < 80\,$km\,s$^{-1}$)}
B stars are line-profile
variables.  We have shown that periodic line-profile variability caused by
surface inhomogeneities or by pulsations can sometimes be mimicked by assuming a
combination of rotational and macroturbulent broadening, but only if these two
parameters are allowed to take on time-dependent values. For some cases, the
interpretation of the line profiles in terms of just $v\sin i\,(t)$ and $v_{\rm
  macro}\,(t)$ is a degradation in quality for the interpretation of the line
broadening compared to full temporal spot or pulsation modelling.

Time series analysis shows that the frequencies found in the $v\sin i\,(t)$ and
$v_{\rm macro}\,(t)$ time series coincide with the spot or pulsation
frequencies { that were already known to occur 
in the moments of the line profiles of the studied stars.} 
Although the basic assumption of dealing with a symmetric line
profile is broken, the F(B)T method remains a powerful tool to deduce $v\sin
i\,(t)$ and subsequently fix that value to deduce $v_{\rm macro}\,(t)$ from a
GOF approach, as already outlined in SDH14.  A caveat is that, when the profile
is particularly asymmetric, saddle points or local extrema occur in the F(B)T
and the interpretation is non-trivial. An appropriate estimate of $v\sin i\,(t)$
can then best be derived by comparing with the F(B)T results
of previous or subsequent
profiles in the time series, where the asymmetry is less during the rotation or
pulsation cycle. Note that considering different spectral lines for the analysis
may also be useful, particularly for spotted stars with chemical
inhomogeneities; pulsations tend to affect all metal lines similarly unless the
magnetic pressure dominates over the gas pressure in the line-forming regions.

Our study shows that one should not rely on a single snapshot spectrum to derive
accurate average values of $v\sin i$ and $v_{\rm macro}$ for main-sequence stars
with slow or moderate rotation, because the effect of spots, and of pulsations
in particular, implies that a relatively large range of these two parameters is
necessary to describe the profile changes over the variability cycles. We have
shown that ignoring the time-variable nature of the line profiles in a snapshot
analysis for $v\sin i$ and $v_{\rm macro}$ may lead to an incorrect
interpretation of the derived values of both quantities. In particular, a high
value of $v_{\rm macro}$ may just occur because the pulsational character of the
star is not taken into account.  The cause of macroturbulent velocities
occurring in line-profile variable O-type stars and B-type supergiants remains
unclear and may be quite different from the case of dwarf stars considered here,
e.g., \citet{SimonDiaz2010} combined with SDH2014.

In a follow-up paper, we plan to present a simulation study where all the
effects that cause a deviation from symmetry of spectral line profiles will be
evaluated in terms of their influence on the derivation of $v\sin i$ from F(B)T
and on $v_{\rm macro}$. The amount of macroturbulent broadening needed to
  mimic spots or pulsations depends not only on the brightness or chemical
  contrasts in the case of spots or on the amplitudes and spherical wavenumbers
  $(\ell,m)$ of the modes in case of pulsations, but also on the inclination
  angle of the star. Moreover, the asymmetry may not only come from spots or
pulsations as discussed here, but also from, e.g., gravity darkening,
deformation from spherical shape due to fast rotation and/or binarity,
reflection effects in close binaries, spectral line blending, etc. In
  addition, the level of the radial and tangential contributions to the
  macroturbulent velocity must be varied from the standard case where they are
  taken equal if the aim is to get a complete picture and solid predictive
  recipes for $v_{\rm macro}$.  This is illustrated in Fig.\,\ref{isoRT} where
  we compare simulated profiles for $v\sin i=15$\,km\,s$^{-1}$ and an isotropic,
  a pure radial, a pure tangential, and an equal-share radial-tangential
  macroturbulent velocity of 20\,km\,s$^{-1}$. It can be seen that the nature of
  the macroturbulent velocity has a large effect on the profile shape. The fact
  that p-mode pulsations typically correspond to a factor 10 to 100 larger
  radial than tangential velocity while it is the other way around for g-mode
  pulsations \citep{Aerts2010}, implies that restricting to $v_{\rm macro}$
  based on an equal share of the radial and tangential contributions, as it was
  mostly done in the literature so far, is not optimal. This issue was already
  addressed in \citet{Aerts2009} where modelling of profiles was done for the
  three extremes (isotropic, radial, tangential) of macroturbulence as also
  shown in Fig.\,\ref{isoRT}.  Only an extensive simulation study in which all
  the phenomena that affect the line profile shape are addressed can lead to a
  full assessment} of the appropriateness, or not, of using only $v\sin i$ and
$v_{\rm macro}$ estimates rather than complete line-profile modelling based on
solid theoretical descriptions, in the evaluation of theoretical stellar
evolution models.


\begin{acknowledgements}
This work was initiated thanks to discussions among the four authors during the
2013 International Francqui Symposium ``What Asteroseismology has to offer to
Astrophysics'' ({\tt http://fys.kuleuven.be/ster/meetings/francqui/francqui})
funded by and organised under the auspices of the Francqui Foundation of
Belgium. This symposium also marked the end of the Advanced Grant: ``Probing
Stellar Physics and Testing Stellar Evolution through Asteroseismology''
(PROSPERITY, {\tt http://www.ster.kuleuven.be/PROSPERITY}) awarded to the first
author by the European Research Council under the European Community's Seventh
Framework Programme (FP7/2007--2013)/ERC grant agreement n$^\circ$227224.  SS-D
acknowledges funding by [a] the Spanish Ministry of Economy and Competitiveness
(MINECO) under the grants AYA2010-21697-C05-04 and Severo Ochoa SEV-2011-0187,
and [b] the Canary Islands Government under grant PID2010119. This research has
made use of the SIMBAD database, operated at CDS, Strasbourg, France.
\end{acknowledgements}


\bibliographystyle{aa}	    
\bibliography{FourierBessel_accepted}		

\end{document}